# AI and Human Approaches to Mathematical Problem Solving

Yang Ding[1,2,3]

[1] University of Edinburgh Business School, Edinburgh EH8 9JS, UK.

[2] Centre for Technomoral Futures, Edinburgh Futures Institute, University of Edinburgh, Edinburgh EH3 9EF, UK.

[3] Manchester Institute of Innovation Research, Alliance Manchester Business School, University of Manchester, Manchester M13 9PL, UK.

Corresponding Author: Yang Ding (yang.ding@ed.ac.uk)

## Abstract

AI systems have begun to report solutions, disproofs, and substantive advances on long-standing mathematical problems, raising questions about whether they approach research in the same way as mathematicians. This study compares public AI research accounts with the human literature on 11 such problems. The human corpus contains 58 papers that directly addressed the same mathematical targets later reported by AI sources as resolved, disproved, or substantially advanced; 31 within-problem comparisons were constructed from these materials. Six validated text-based measures capture problem resolution, method articulation, uncertainty and boundary specification, successor-question generation, generality, and cross-disciplinary integration. AI accounts place greater emphasis on resolving the focal problem and connecting ideas across fields. Human papers devote significantly more attention to explaining methods, specifying assumptions and limitations, and identifying questions for subsequent research. No precise difference is detected in generality. The estimated directions remain unchanged when each mathematical problem is removed in turn. The findings reveal two distinct research profiles: AI accounts concentrate on closing and recombining problems, whereas mathematical papers more extensively document the procedures, limits, and research opportunities through which results become cumulative knowledge. Evaluating research AI therefore requires attention to the organization of inquiry, not only whether a target is solved.

**Keywords:** Artificial intelligence; Mathematical problem solving; Human–AI comparison; Research practices; Knowledge production; Computational text analysis

**JEL classification:** C55; D83; O31; O33

# 1. Introduction

Research mathematics does more than decide whether statements are true. It develops human understanding through conjectures, proofs, reusable methods, domains of validity and questions that organize subsequent inquiry. Rapid advances in research-level AI have therefore prompted concern not only about technical performance, but also about the future organization of mathematical work, the training of new mathematicians and the community's capacity to build and transmit understanding. In a joint declaration signed by 25 Fields Medalists, including Terence Tao, these concerns were framed in relation to mathematical understanding, education, community and the discipline's longer-term purposes (Tao 2026). This broader concern motivates an evaluation of mathematical AI that asks not only whether a target is solved, but also which practices supporting cumulative research are preserved in its public outputs.

Research on AI in science distinguishes technical capability from changes in knowledge production. Bianchini et al. (2022) describe AI as an emerging method of invention that can reshape discovery and the organization of science. In mathematics, machine learning has guided human conjecture formation (Davies et al. 2021), generated conjectures (Raayoni et al. 2021), searched for new constructions (Romera-Paredes et al. 2024) and supported formal or semi-autonomous problem solving (Feng et al. 2026; Trinh et al. 2024). He (2024) groups these developments into bottom-up, top-down and meta-mathematical approaches. Commelin et al. (2026) and Nature Machine Intelligence (2026) broaden the discussion to verification, research practice and the organization of mathematical work. Together, this literature motivates examining AI as part of a knowledge-production system rather than only as a solver of isolated tasks.

Capability gains alone do not show how AI-generated results enter cumulative mathematical research. Tao (2026) argues that using outstanding problems primarily as AI benchmarks can conflict with the broader aims of mathematical inquiry. This concern makes it important to measure more than computational speed or stylistic similarity. We examine the observable exploration profile of each text: the attention it gives to problem closure, method accumulation, boundary specification, agenda formation and knowledge integration.

This paper compares AI and human research texts organized around the same mathematical problems. It asks a deliberately observable and limited question: how do the two types of text differ across six research practices? The analysis does not recover private cognition, certify mathematical correctness or estimate a causal effect of AI. Texts are treated as public records of how mathematical work is presented to a research community.

Research-level AI results now appear before the relevant communities have settled how to describe and compare their contribution to cumulative inquiry. A closure-only evaluation can overlook whether a text explains reusable methods, identifies limits or opens subsequent lines of work. A multidimensional comparison provides evidence about these components without presuming that any one of them defines mathematical quality.

The paper makes three contributions. First, it operationalizes six observable research practices that connect mathematical problem solving to cumulative knowledge production. Second, it compares AI and human texts within the same target problem and gives each problem equal inferential weight, limiting confounding from differences in subject matter and literature size. Third, it shows which contrasts persist across text representations, matching rules and problem composition, clarifying where the evidence is stable and where it remains sensitive. This framework links research-level mathematical AI to management research on the organization and evaluation of scientific discovery.

## 2. Data

### 2.1 Sources and unit of observation

The mathematical problem is the unit of analysis. The full retrieval and measurement-development corpus contains 36 AI problem documents and 980 human abstracts across 23 problems; it supports problem identification, recall and construct validation but is not the final effect-estimation sample.

The confirmatory same-problem sample covers 11 mathematical problems, 58 human comparator papers selected by the target-linkage audit and 31 AI-human matched pairs. A separate set of 497 parseable full texts is used for abstract-full-text representation checks. The AI corpus draws on publicly available research documents from several source clusters, including a Gemini-assisted study of the Erdős problems (Feng et al. 2026), a Rethlas-based investigation of open problems in commutative algebra (Jiang et al. 2026), and OpenAI's collection of ten advances in mathematics and theoretical computer science (OpenAI 2026).

## 2.2 Mathematical problems represented

The analytical sample contains 11 target problems, conjectures or sustained research challenges documented in primary sources. The entries differ in scope and evidential status. Some sources report affirmative or negative resolutions of specified propositions, whereas others report advances on narrower theoretical targets. The unit-distance entry concerns a particular asymptotic conjecture, the Navier-Stokes entry concerns the smoothly forced Clay C/D formulation, and the coding and permanent entries are treated as advances rather than resolutions of their fields' broader open problems. Table 1 records the target and the status claimed by each source.

**Table 1. Mathematical problems and AI-result roles in the analytical sample.**

| Mathematical problem | Field | Analytical role of AI result |
|---|---|---|
| Forced Navier-Stokes blow-up | PDE / fluid mechanics | Specified-form resolution |
| Unforced Euler blow-up | PDE / fluid mechanics | Resolution |
| Planar unit-distance conjecture | Discrete geometry | Disproof |
| Erdos Problem 728 | Number theory | Affirmative resolution |
| Cohn-Elkies asymptotic rate | Discrete geometry | Stated-conjecture resolution |
| Binary and spherical code bounds | Coding theory | Substantive progress |
| Permanent lower bounds | Algebraic complexity | Substantive progress |
| Quantum parallel repetition | Quantum information | Affirmative resolution |
| Closest vector hardness | Theoretical computer science | Fixed-target resolution |
| Anderson Problem 8a | Commutative algebra | Counterexample / disproof |
| Eisenbud-Schreyer Question 6.1 | Commutative algebra | Counterexample / disproof |

### 2.3 Identification, cleaning and comparator audit

Candidate records are retrieved using canonical names, aliases, mathematical objects and problem-specific anchors. Cleaning removes generic-name matches, disciplinary false positives, non-research items and records not directed at the target statement. The audit distinguishes complete resolution, negative resolution, resolution of a specified formulation and substantive progress. Human comparator records are retained when the observed text links a result-oriented predicate to the target statement under the prespecified audit rules. This procedure identifies textual comparability; it is not presented as independent certification of mathematical correctness. **Appendix A** documents the data sources, retrieval procedures and construction of the problem-level corpus. **Appendix B** reports the cleaning decisions, eligibility criteria and target-linkage audit rules used to select the human comparator papers.

## 3. Methods

### 3.1 Six-dimensional research-practice framework

The six indicators describe how a text presents research. They capture a claim to close the target problem, the articulation of a reusable method, the statement of assumptions or failure boundaries, the generation of successor questions, extension beyond the focal case and explicit integration across mathematical fields. The first five are semantic orientations estimated from word and phrase distributions. Cross-disciplinary integration is estimated separately from relational features connecting claims, objects, methods and fields. The distinction follows the constructs' informational content rather than the observed AI-human effects.

Figure 1 provides an integrated map of how heterogeneous research texts are converted into comparable problem-level profiles. Its organizing principle is separation: corpus construction is distinguished from confirmatory comparison, construct definition from estimator selection, and document-level measurement from problem-level inference. This architecture is designed to prevent retrieval breadth, variation in textual representation and uneven publication volume from implicitly redefining the object of comparison. The four panels should therefore be read as a connected sequence of design decisions that preserves a consistent comparison between public AI accounts and mathematical papers.

Measurement and inference architecture
A Data architecture
Development corpus
23 problems
36 AI documents
980 human abstracts
Target-linkage audit
Result predicate
Target anchor
Affirmative relation
Confirmatory sample
11 problems
58 comparator papers
31 matched comparisons
Full-text check
497 human papers
Abstract and full text
Original-text check
Complete AI texts
5 mathematical problems
B Constructs and measurement
Closure
Problem resolution
Reuse
Method articulation
Limits
Uncertainty and boundaries
Agenda
Successor questions
Scope
Generality
Bridges
Cross-disciplinary integration
Semantic-distribution measurement
Five research orientations
Word and phrase distributions
Independent known-group validation
Claim-relation measurement
C Within-problem estimation
1 Match
Up to five human texts Nearest log token length
2 Standardize
Common definitions Comparable score scale
3 Aggregate
One contrast per problem Comparator papers averaged
4 Estimate
Equal problem weights Problem-cluster bootstrap
D Uncertainty and robustness
Problem-cluster bootstrap
95% intervals
Problem-level dependence
Leave one problem out
11 omission estimates
All six signs preserved
Abstract and full text
497 human papers
Rank correlations 0.637-0.831
Original and structured text
Five-problem check
All six directions retained

**Figure 1. Measurement and inference architecture.** A, Separation of the measurement-development corpus, target-linkage audit, confirmatory same-problem sample, and text-representation checks. B, Six research-practice constructs and their construct-aligned measurement approaches. Five constructs are estimated from semantic distributions, whereas cross-disciplinary integration is estimated from claim relations. C, Within-problem matching, standardization, aggregation, equal problem weighting, and problem-cluster bootstrap inference. D, Leave-one-problem-out, abstract–full-text, and original–structured-text robustness checks.

## 3.2 Measurement models, matching and problem-level inference

The first five constructs use independently estimated, prespecified semantic models. For document i and construct k, a TF-IDF representation of word unigrams and bigrams enters an L2-regularized logistic regression with balanced class weights (C=0.7). Each model is trained on an external known-group corpus. Equation (1) gives the probability assigned to the high-construct group. Resolution contrasts theorem-oriented papers with reviews; method articulation contrasts methods papers with result announcements; uncertainty and boundaries contrasts limitation-rich with ordinary research texts; successor questions contrasts open-problem with ordinary research texts; and generality contrasts generalizations with special cases.

$$s_{ik} = \Lambda(\hat{\alpha}_k + \phi(x_i)' \hat{\beta}_k), \quad k \in \{\text{resolution}, \text{method}, \text{boundary}, \text{successor questions}, \text{generality}\} \tag{1}$$

where $\Lambda(z) = 1/1 + e^{-z}$ , $\phi(x_i)$ contains sublinearly scaled TF-IDF features and the fitted coefficients use the complete external training corpus. Five-fold prediction is used only to estimate validation performance: DOI hashes assign folds, so each validation document is scored by a model that did not observe it. The resulting out-of-fold AUCs range from 0.926 to 0.990.

Cross-disciplinary integration uses a separate supervised model because field-word co-occurrence does not establish integration. The feature map $g(x_i)$ contains 28 normalized lexical and relational features, including claim-object, method-operation, condition, scope, generalization, field-bridge and mathematical-field entropy measures. A standardized histogram-based gradient-boosting classifier maps these features to the probability that a document belongs to the cross-disciplinary known group:

$$c_i = \Pr(Y_i^c = 1 | \ g(x_i); \hat{h}) \tag{2}$$

where $\hat{h}$ denotes the fitted gradient-boosting function. Five-fold out-of-fold prediction on the external known groups yields an AUC of 0.849. The fitted model then scores the focal AI and human texts. **Appendix C** reports the feature families, model settings and validation results. It also records the internal implementation labels used in the reproducible workflow.

To assess sensitivity to text representation, we re-estimate all six constructs in the five-problem subset for which complete original AI texts are available. This complementary analysis assigns estimators according to the observable structure of each construct. Latent-semantic logistic models measure problem resolution, method articulation, uncertainty and boundaries, and successor-question generation, whereas claim-relation models measure generality and cross-disciplinary integration because these constructs depend more directly on scope and field-linkage relations. Both model families are trained and evaluated using independent external known groups.

Within each mathematical problem, AI document $i$ is matched to as many as five human texts $h$ with the smallest absolute difference in log token count, $d_{ih} = |\log L_i - \log L_h|$. Matching addresses the mechanical association between document length and opportunities to express a construct. The final design contains 31 matched comparisons across 11 problems and draws on 58 available human comparator papers.

$$\hat{\Delta}_k = P^{-1} \sum_{p=1}^{P} \frac{\bar{s}_{pk}^{\mathrm{AI}} - \bar{s}_{pk}^{\mathrm{H}}}{\sigma_{pk}^{\mathrm{H}}} \tag{3}$$

where P=11, the bars denote mean scores within problem $p$ and $\sigma_{pk}^{\mathrm{H}}$ is the human-paper standard deviation for construct k within that problem. Papers are averaged before problems receive equal weight. The 95% intervals resample problems, and leave-one-problem-out estimates repeat Equation (3) after removing each problem in turn.

# 4. Results

In the same-problem design, AI texts score 1.01 human within-problem standard deviations higher in problem-resolution orientation (95% problem-bootstrap interval 0.05 to 2.08) and 0.86 higher in cross-disciplinary integration (0.17 to 1.46). Human texts score 1.18 standard deviations higher in method articulation (AI-human estimate -1.18, interval -1.85 to -0.53), 0.72 higher in uncertainty and boundaries (-0.72, -1.27 to -0.19), and 1.09 higher in successor-question generation (-1.09, -1.90 to -0.39). The generality estimate is -0.12 (-0.50 to 0.23). Five intervals exclude zero; generality does not. Figure 2 places these aggregate differences alongside their stability across problem omissions, the underlying problem-level variation and the distribution of human comparator evidence.

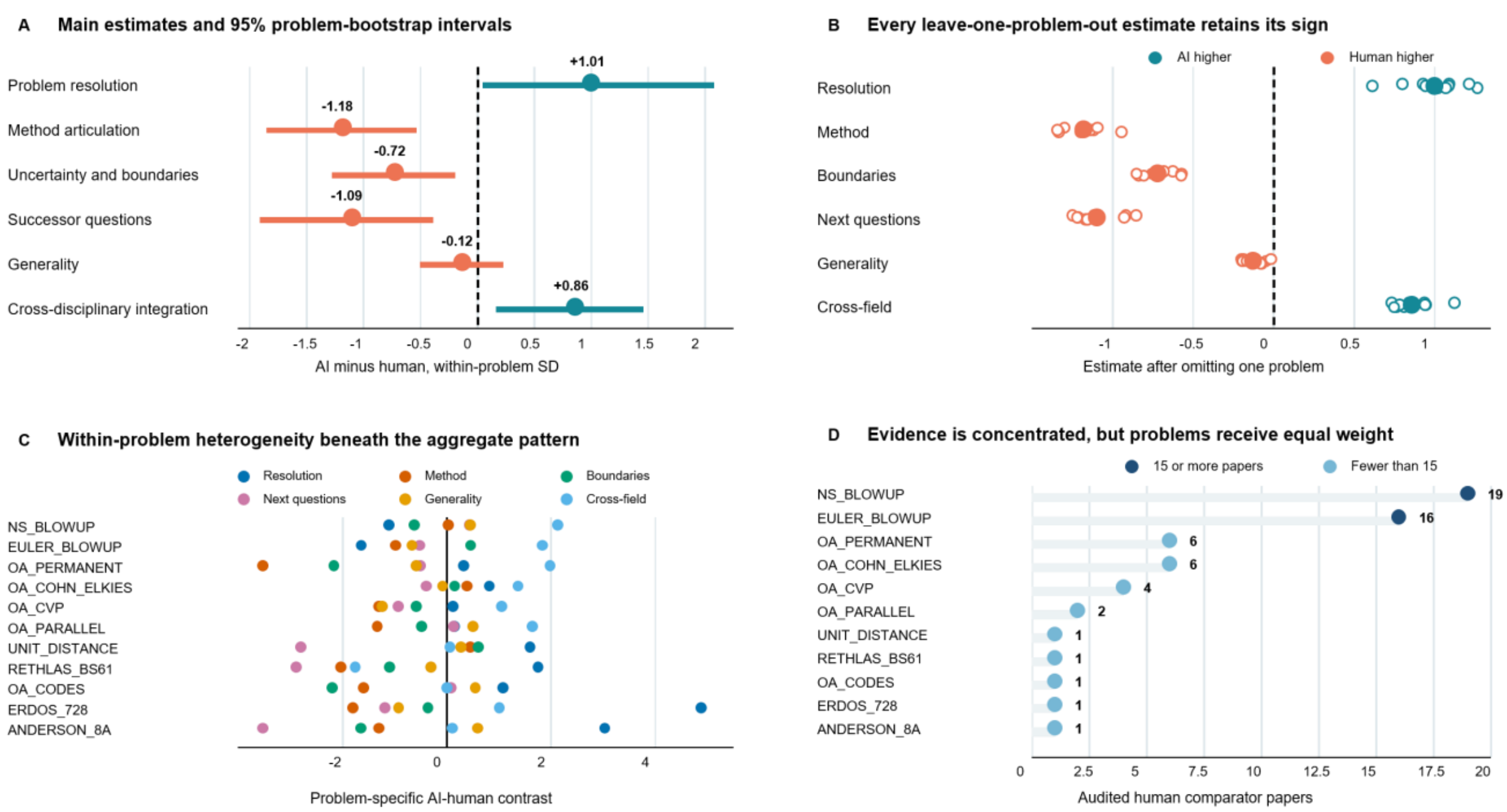

**Figure 2. Main AI-human differences, leave-one-problem-out stability, problem-level heterogeneity and comparator composition.** A, Equal-weight same-problem estimates with 95% problem-bootstrap intervals; positive values indicate greater emphasis in AI texts. B, Estimates after separately omitting each of the 11 problems; filled and open points represent full-sample and omission estimates, respectively. All signs are preserved. C, Problem-specific contrasts, with colors identifying the six constructs. D, Distribution of the 58 human comparator papers; dark and light blue indicate problems represented by at least 15 and fewer than 15 papers, respectively. Problems receive equal weight in aggregate estimation.

Leaving out each of the 11 mathematical problems in turn preserves the direction of all six aggregate estimates (Figure. 2B). Problem resolution and cross-disciplinary integration remain positive, whereas method articulation, uncertainty and boundaries, successor-question generation and generality remain negative. The complete omission estimates reported in **Appendix D** show that no single problem determines the aggregate pattern. Generality nevertheless remains close to zero and therefore exhibits directional, rather than statistical, stability.

**Appendix E** examines whether the findings depend on representing AI research through structured summaries. In the five problems for which complete original AI texts are available, the construct-aligned alternative measurement reproduces the direction of all six main estimates. Each alternative measure also achieves an external known-group AUC above 0.80.

**Appendix F** evaluates whether the abstract-based measures preserve the broader ordering observed in full-text representations. Across 497 human papers, scores derived from abstracts are positively correlated with those derived from equal-length full-text windows for all six constructs, with Spearman correlations ranging from 0.637 to 0.831. These results support broad rank correspondence between the two representations.

Sensitivity to comparator selection and matching specification is assessed in **Appendix G**. The estimated directions of problem resolution, method articulation, uncertainty and boundaries, successor-question generation, and cross-disciplinary integration remain unchanged across the alternative specifications. The generality estimate retains the main direction under the closest-75% specification but changes sign under three other specifications, with the estimates remaining close to zero. This sensitivity is consistent with the main analysis, which identifies no statistically precise AI-human difference in generality. The alternative specifications therefore reinforce the five directional findings while supporting a null interpretation for generality.

# 5. Discussion

The estimates do not produce a single ordering of AI and human mathematical research. The sampled AI texts place more emphasis on closure claims and cross-field relations. The sampled human texts place more emphasis on methods, boundary conditions and successor questions. These are differences in public research presentation, not direct observations of private cognition. Their multidimensional form is consistent with accounts of AI as a component of changing mathematical practice rather than a uniform substitute for mathematicians (Commelin et al. 2026; He 2024).

The comparison also clarifies Tao (2026) concern about treating outstanding problems primarily as AI benchmarks. A stronger closure orientation does not imply greater emphasis on every practice that supports cumulative inquiry. Method articulation, boundary specification and successor-question generation remain empirically separate dimensions. Evaluating research-level

AI with a single success criterion can therefore conceal differences in how results are prepared for interpretation and further use.

These differences also have implications for how AI-generated mathematical work enters the research system. Public accounts of mathematical results provide the materials through which claims are scrutinized, methods are reused and research agendas are extended. The sampled AI texts represent an early and institutionally concentrated stage of development, when conventions for documenting, evaluating and disseminating AI-assisted research are still taking shape. The UK AI Security Institute's *Frontier AI Trends Report* identifies hypothesis generation, experimental design, outcome prediction, research pace and benefits across users as important dimensions in the evaluation of AI for science (AI Security Institute 2025). Our analysis points to a related institutional requirement: AI-generated results need forms of presentation that make their procedures, domains of validity and implications for subsequent inquiry accessible to mathematical communities. The development of research-level AI therefore involves not only advances in model capability, but also publication and evaluation arrangements through which individual outputs can contribute to cumulative mathematical knowledge.

Future research can examine whether AI-generated methods are reused, whether proposed successor questions attract subsequent work and whether human-AI collaborations produce different research-practice profiles. Such evidence would connect the presentation differences documented here to later knowledge accumulation.

# Appendix A. Problem identification and retrieval

Each problem receives a canonical name, aliases, mathematical statement, problem-specific anchors, family, AI source and claim status. Retrieval takes the union of name queries, anchor co-occurrence queries and semantic recall, then retains records that connect to the target statement in title or abstract. Table A1 provides the corresponding audit trail: the precise scope of each target, and the boundary of what the source actually claims.

**Table A1. Target definitions, claim boundaries and source verification.**

| ID | Precise target statement | Boundary of source claim |
|---|---|---|
| NS_BLOWUP | Forced 3D incompressible Navier-Stokes finite-time blow-up (Clay statements C/D) | Resolution is restricted to the forced C/D formulation |
| EULER_BLOWUP | Unforced 3D incompressible Euler finite-time blow-up | Analytic counterexample with Lean formalization |
| UNIT_DISTANCE | Erdos's $n^{1+o(1)}$ conjecture for planar unit distances | Disproof of the conjectured asymptotic upper behaviour |
| ERDOS_728 | A strengthened non-trivial formulation of Erdős Problem 728 on factorial divisibility | Affirmative answer to a strengthened formulation interpreted as capturing the intended question; the original statement is ambiguous |
| OA_COHN_ELKIES | Conjectured high-dimensional rate of the Cohn-Elkies sphere-packing linear program | Resolution of the stated asymptotic-rate conjecture |
| OA_CODES | High-dimensional linear-programming bounds for unrestricted binary and spherical codes | Strict exponent improvements; not a complete closure claim |
| OA_PERMANENT | General division-free arithmetic-circuit and arithmetic-formula lower bounds for the permanent | Permanent-specific superquadratic lower bounds for division-free circuits and order-$n^4/\log n$ formula lower bounds; VP versus VNP remains open |
| OA_PARALLEL | Quantum parallel repetition for finite two-player one-round entangled games | Affirmative resolution for the stated finite-game and finite-dimensional strategy setting |
| OA_CVP | $n^{1/400}$-factor NP-hardness of the Euclidean closest vector problem | Resolution of the stated $n^{1/400}$ -approximation hardness target |
| ANDERSON_8A | Anderson's weak quasi-completeness conjecture (Problem 8a) | Negative resolution by counterexample |
| RETHLAS_BS61 | Eisenbud-Schreyer Question 6.1 on pure rays for the Heisenberg Lie algebra | Negative resolution of the stated realization question |

# Appendix B. Comparator audit and sample construction

The comparator audit is a reproducible textual screen. It requires a result-oriented predicate, a target anchor and an affirmative relation in the same sentence, while flagging negation, still-open language, partial results, conditional claims, special cases, averaged settings and forced formulations. Fifty-eight of 440 broad candidates satisfy the prespecified textual rule across 11 problems. Table B1 shows that the resulting evidence base is uneven: Euler and Navier-Stokes supply most comparator records, whereas five problems contribute one record each.

**Table B1. Human comparator papers passing the textual target-linkage audit.**

| Problem | Comparator papers |
|---|---|
| ANDERSON_8A | 1 |
| ERDOS_728 | 1 |
| EULER_BLOWUP | 16 |
| NS_BLOWUP | 19 |
| OA_CODES | 1 |
| OA_COHN_ELKIES | 6 |
| OA_CVP | 4 |
| OA_PARALLEL | 2 |
| OA_PERMANENT | 6 |
| RETHLAS_BS61 | 1 |
| UNIT_DISTANCE | 1 |

# Appendix C. Measurement models and construct validity

The five semantic constructs use separate TF-IDF logistic regressions trained on independent known groups. Word unigrams and bigrams use sublinear term frequency, a maximum of 80,000 features, minimum document frequency of two and maximum document frequency of 0.98. Logistic regressions use balanced class weights, L2 regularization and C=0.7. Equation (A1) defines the focal-document score. Cross-disciplinary integration uses the 28-feature relational map g(x) and histogram-based gradient boosting specified in Equation (A2), with 180 iterations, at most 12 leaf nodes, learning rate 0.055 and L2 regularization of 2.0.

$$s_{i,k} = \Lambda(\alpha_k + \phi(x_i)^T \beta_k^{-f(i)}) \tag{A1}$$

where $s_{i,k}$ denotes the assigned score of focal document $i$ for construct $k \in$ {resolution,method,boundary,successor questions,generality}; $\Lambda(z) = (1 + e^{-z})^{-1}$ represents the standard logistic sigmoid link function; $\alpha_k$ is the construct-specific intercept; $\phi(x_i)$ is the feature vector of sublinearly scaled TF-IDF metrics for document $i$; and $\beta_k^{-f(i)}$ denotes the fitted coefficient vector for construct $k$ estimated from the external training corpus with the fold $f(i)$ containing document $i$ held out (thus guaranteeing out-of-fold scoring).

$$c_i = \Pr(Y_i^c = 1 \mid g(x_i); \hat{h}) \tag{A2}$$

In Equation (A2), $c_i$ represents the cross-disciplinary integration probability for document $i$; $Y_i^c \in \{0,1\}$ is the binary group membership indicator; $g(x_i)$ is a 28-dimensional vector of normalized lexical and relational features capturing structural claim-object, method-operation, scope, and mathematical-field entropy measures; and $\hat{h}$ denotes the fitted histogram-based gradient-boosting classifier.

DOI hashes assign the five validation folds. The out-of-fold AUCs evaluate discrimination in the external known groups; the focal AI-human scores come from models refitted on the complete external corpus. Table C1 reports five-fold out-of-fold validation.

**Table C1. Independent known-group validation of the six measurement models.**

| Construct | Estimator family | Five-fold AUC | Validation interpretation |
|---|---|---|---|
| Problem resolution | | 0.990 | Distinguishes theorem-oriented from review texts |
| Method articulation | | 0.984 | Distinguishes methods from result announcements |
| Uncertainty and boundaries | Semantic distribution | 0.926 | Distinguishes limitation-rich texts |
| Successor questions | | 0.987 | Distinguishes open-problem texts |
| Generality | | 0.928 | Distinguishes generalizations from special cases |
| Cross-disciplinary integration | Claim-relation network | 0.849 | Distinguishes cross-field from pure-field texts |

# Appendix D. Leave-one-problem-out analysis

Each specification omits one of 11 problems and recomputes the equal-weight mean over the remaining ten, as defined in Equation (A3).

$$\hat{\Delta}_k^{(-q)} = (P-1)^{-1} \sum_{p \neq q} \frac{\overline{s}_{pk}^{\mathrm{AI}} - \overline{s}_{pk}^{\mathrm{H}}}{\sigma_{pk}^{\mathrm{H}}} \tag{A3}$$

In Equations (3) and (A3), $\hat{\Delta}_k$ is the aggregate standardized mean effect size for construct $k$ across all $P$ problems (where $P = 11$); $\hat{\Delta}_k^{(-q)}$ is the leave-one-problem-out effect size calculated after omitting problem $q$; $\bar{s}_{pk}^{\mathrm{AI}}$ and $\bar{s}_{pk}^{\mathrm{H}}$ denote the mean scores of AI and human texts within problem $p$ on construct $k$, respectively; $\sigma_{pk}^{\mathrm{H}}$ is the human-paper standard deviation for construct $k$ within problem $p$; and $P-1$ accounts for the sample size adjustment following problem omission.

Table D1 contains the minimum and maximum across all omissions. None of the six ranges crosses zero. Generality comes closest, with a maximum of -0.013; this establishes sign stability in the leave-one-out exercise.

**Table D1. Leave-one-problem-out effect ranges.**

| Construct | Minimum LOO effect | Maximum LOO effect | Sign always preserved |
|---|---|---|---|
| Uncertainty and boundaries | -0.849 | -0.569 | Yes |
| Cross-disciplinary integration | 0.736 | 1.125 | Yes |
| Generality | -0.197 | -0.013 | Yes |
| Method articulation | -1.340 | -0.942 | Yes |
| Successor questions | -1.248 | -0.852 | Yes |
| Resolution | 0.617 | 1.269 | Yes |

# Appendix E. Construct-aligned hybrid measurement

This appendix remeasures all six research-practice constructs using the available complete original AI texts to assess whether the main results depend on structured summaries. Complete original texts can be matched to human papers for five mathematical problems. The exercise therefore evaluates directional and representation sensitivity rather than replacing the 11-problem main analysis.

Different constructs require different estimators because the six research practices have different observable structures in text. Applying one bag-of-words or semantic model to every construct would be formally simple, but it would assume that all research practices become visible through the same linguistic mechanism and could therefore create construct-estimator mismatch.

Problem resolution, method articulation, uncertainty and boundaries, and successor questions are diffuse discourse functions. Their evidence is distributed across sentences and may appear through semantically related formulations that do not share exact keywords. Latent-semantic logistic models compress correlated expressions into lower-dimensional semantic components and are therefore suited to these document-level rhetorical orientations. Generality and cross-disciplinary integration instead depend on explicit relations among elements within a claim. Generality requires a scope-extending predicate to be linked to the focal result; cross-disciplinary integration requires concepts from distinct fields to participate in the same substantive assertion. Mere vocabulary co-occurrence cannot establish these relations, so both constructs use claim-relation models that encode predicate-target, scope and field-linkage features.

All six estimates have the same direction as the main analysis, and every external AUC exceeds 0.80, as shown in Table E1. The generality interval excludes zero; the other five intervals span zero with only five problem clusters. The result is therefore a complete directional replication. Because the construct-model combination was formulated after preliminary representation checks, this appendix treats it as an exploratory robustness analysis.

**Table E1. Original-text results from the construct-aligned hybrid measurement.**

| Construct | Model | External AUC | Effect | 95% interval | Same direction as main |
|---|---|---|---|---|---|
| Problem resolution | Latent-semantic Logit | 0.992 | +0.050 | [-0.243, 0.418] | Yes |
| Method articulation | Latent-semantic Logit | 0.987 | -0.534 | [-2.064, 1.201] | Yes |
| Uncertainty and boundaries | Latent-semantic Logit | 0.927 | -0.322 | [-1.438, 0.972] | Yes |
| Successor questions | Latent-semantic Logit | 0.976 | -0.114 | [-0.778, 0.716] | Yes |
| Generality | Claim-relation model | 0.891 | -0.899 | [-1.567, -0.261] | Yes |
| Cross-disciplinary integration | Claim-relation model | 0.849 | +0.584 | [-0.414, 1.552] | Yes |

# Appendix F. Abstract and full-text representations

Table F1 shows statistically detectable positive correlations for all six constructs. Their magnitudes support rank correspondence, with the weakest agreement for cross-disciplinary integration. Across 497 human papers, abstract scores correlate with scores from equal-length full-text windows at 0.637 to 0.831 (Spearman's rho). This supports similarity in broad document rankings, not equality between abstracts and full texts. The lower correlation for cross-disciplinary integration is consistent with relational evidence appearing outside abstracts.

**Table F1. Abstract and equal-length full-text score correlations.**

| Construct | Papers | Abstract-equal-length full-text rho | *p* |
|---|---|---|---|
| Problem resolution | 497 | 0.831 | <0.001 |
| Method articulation | 497 | 0.767 | <0.001 |
| Uncertainty and boundaries | 497 | 0.781 | <0.001 |
| Successor-question generation | 497 | 0.818 | <0.001 |
| Generality | 497 | 0.741 | <0.001 |
| Cross-disciplinary integration | 497 | 0.637 | <0.001 |

# Appendix G. Matching, influence and uncertainty

The inverse-distance weighting follows Equation (A4):

$$\omega_{ih} = \frac{(d_{ih} + \epsilon)^{-1}}{\sum_{h \in H_i} (d_{ih} + \epsilon)^{-1}} \tag{A4}$$

where $\omega_{ih}$ denotes the normalized inverse-distance weight assigned to human paper $h$ matched with focal AI document $i$; $d_{ih} = |\log L_i - \log L_h|$ represents the absolute log token-count distance between AI text $i$ and human text $h$; $\epsilon > 0$ is a small smoothing constant to avoid division by zero; and $H_i$ denotes the set of matched human comparator papers for focal AI document $i$.

Table G1 shows stable directions for problem resolution, method articulation, uncertainty and boundaries, successor questions and cross-disciplinary integration. Generality changes sign in three alternatives, consistent with its imprecise main estimate.

**Table G1. Matching-specification sensitivity.**

| Specification | Construct | Problems | Effect | Main sign |
|---|---|---|---|---|
| all comparator pairs | Uncertainty and boundaries | 11 | -0.72 | Yes |
| all comparator pairs | Cross-disciplinary integration | 11 | 0.86 | Yes |
| all comparator pairs | Generality | 11 | -0.12 | Yes |
| all comparator pairs | Method articulation | 11 | -1.18 | Yes |
| all comparator pairs | Successor-question generation | 11 | -1.09 | Yes |
| all comparator pairs | Problem resolution | 11 | 1.01 | Yes |
| best one-to-one | Uncertainty and boundaries | 11 | -0.86 | Yes |
| best one-to-one | Cross-disciplinary integration | 11 | 0.99 | Yes |
| best one-to-one | Generality | 11 | 0.27 | No |
| best one-to-one | Method articulation | 11 | -1.05 | Yes |
| best one-to-one | Successor-question generation | 11 | -1.20 | Yes |
| best one-to-one | Problem resolution | 11 | 0.92 | Yes |
| closest 50% | Uncertainty and boundaries | 11 | -0.94 | Yes |
| closest 50% | Cross-disciplinary integration | 11 | 0.96 | Yes |
| closest 50% | Generality | 11 | 0.02 | No |
| closest 50% | Method articulation | 11 | -1.17 | Yes |
| closest 50% | Successor-question generation | 11 | -1.08 | Yes |
| closest 50% | Problem resolution | 11 | 0.90 | Yes |
| closest 75% | Uncertainty and boundaries | 11 | -0.86 | Yes |
| closest 75% | Cross-disciplinary integration | 11 | 0.79 | Yes |
| closest 75% | Generality | 11 | -0.11 | Yes |
| closest 75% | Method articulation | 11 | -1.24 | Yes |
| closest 75% | Successor-question generation | 11 | -1.11 | Yes |
| closest 75% | Problem resolution | 11 | 1.01 | Yes |
| inverse-distance weighted | Uncertainty and boundaries | 11 | -0.81 | Yes |
| inverse-distance weighted | Cross-disciplinary integration | 11 | 0.91 | Yes |
| inverse-distance weighted | Generality | 11 | 0.06 | No |
| inverse-distance weighted | Method articulation | 11 | -1.19 | Yes |
| inverse-distance weighted | Successor-question generation | 11 | -1.06 | Yes |
| inverse-distance weighted | Problem resolution | 11 | 0.93 | Yes |